\documentclass[floatfix,preprintnumbers,amsmath,amssymb,prl,nofootinbib]{revtex4}
\usepackage{graphicx}
\def\pmb#1{\setbox0=\hbox{#1}
\kern-.025em\copy0\kern-\wd0
\kern.05em\copy0\kern-\wd0
\kern-.025em\raise.0433em\box0}

\newcommand{\etal}{{\it et al. }}

\begin{document}

\textheight=25cm

\setlength{\oddsidemargin}{-1.3 true cm}
\setlength{\evensidemargin}{-1.5 true cm}

\setlength{\textheight}{25 true cm}
\setlength{\textwidth}{17.0 true cm}

\setlength{\footskip}{1.8 True cm}
\setlength{\headheight}{0. true cm}
\setlength{\headsep}{1.2 true cm}
\setlength{\topmargin}{-2  true cm}
\setlength{\voffset}{-1.5 true cm}

\large

\title{\Large Reply to:  ``Comment on `Spurious fixed points in frustrated magnets,'cond-mat/0609285''}

\large

\author{\Large B. Delamotte$^{1}$, Yu. Holovatch$^{2,3}$, D. Ivaneyko$^{4}$, D. Mouhanna$^{1}$
 and M. Tissier$^{1}$}

\address{\large $^{1}$ LPTMC, CNRS-UMR 7600, Universit\'e Pierre
et Marie Curie, 75252 Paris C\'edex 05, France}

\address{\large $^{2}$
Institute   for Condensed  Matter   Physics  of the National Acad. Sci.  of Ukraine, 
UA--79011 Lviv, Ukraine}

\address{\large $^{3}$
Institut  f\"ur  Theoretische Physik, Johannes  Kepler Universit\"at Linz,
A-4040 Linz, Austria}

\address{\large $^{4}$
Ivan Franko National University of Lviv, UA--79005 Lviv, Ukraine}

\begin{abstract}

\large

The Comment      of A.  Pelissetto and E.        Vicari (cond-mat/0610113) 
on our article (cond-mat/0609285) is based on  misunderstandings of this article as
well  as  on unfounded  implicit  assumptions.   We  clarify here  the
controversial points  and show that, contrary to  what is  asserted by
these authors, our paper is free of any  contradiction and agrees with
all  well-established theoretical and  experimental results.  Also, we
maintain that  our  work  reveals  pathologies in   the (treatment of)
perturbative approaches performed  at fixed dimensions. In particular,
we emphasize that  the  perturbative approaches to frustrated  magnets
performed either within the minimal substraction  scheme without $\epsilon$-expansion or in
the   massive scheme at zero    momentum exhibit {\sl spurious}  fixed
points and, thus,  do  not describe correctly  the behaviour  of these
systems in three dimensions.

\end{abstract}

\maketitle

\newpage


\section{\large I- Introduction}

Before answering in detail to the technical points raised in Comment
\cite{pelissetto06} we would like to point out that the authors of
\cite{pelissetto06} very often quote our results in a biased way, as if our article 
\cite{delamotte06} was written to emphasize the differences between the 
{\sl  perturbative} fixed dimension (FD) approaches   and the {\sl non
perturbative} renormalization  group (NPRG) approach (named functional
renormalization  group  (FRG)   approach  in \cite{pelissetto06}),  an
approach that has been employed by some  of us in previous articles to
investigate the physics of frustrated magnets
\cite{tissier00b,tissier00,tissier01,delamotte03}. 

This is {\sl absolutely} not the case.  Our aim, in our article
\cite{delamotte06},  was to shed light on the discrepancy between the
different {\sl  perturbative} approaches, namely the $\epsilon$-expansion and
the FD approaches.  The NPRG results were  quoted just as side remarks
--- to show the  agreement between NPRG  and $\epsilon$-expansion --- and   we
have only dealt in \cite{delamotte06} with {\sl perturbative} methods.

\begin{figure}[htbp] 
\vspace{-0cm}
\hspace{2cm}

\includegraphics[width=0.6\linewidth,origin=tl]{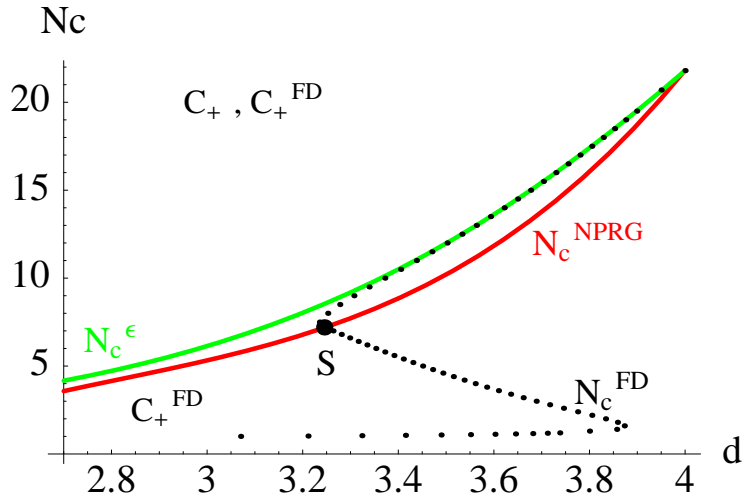}\hfill%
\caption{Curves $N_c(d)$ obtained within the  $\overline{\hbox{MS}}$ with an $\epsilon$-expansion ($N_c^\epsilon$), the  $\overline{\hbox{MS}}$ scheme   without   $\epsilon$-expansion  ($N_c^{\text{FD}}$)  and the   NPRG
 approach    ($N_c^{\text{NPRG}}$).     The    part  of    the   curve
 $N_c^{\text{FD}}$     below S     corresponds    to  a    regime   of
 non-Borel-summability.}
\label{courbes_ncd} 
\end{figure}

 The    existence  of  such       a  discrepancy  is      evident from
 Fig.\ref{courbes_ncd} \footnote{\large As a  precaution we emphasize,  as in
 \cite{delamotte06}, that a large part  of the curve $N_c^{\text{FD}}$
 obtained  within the FD  approach, typically that below the ``turning
 point'' S,  corresponds to fixed points (FPs) that  are situated out of
 the Borel-summability region.  Thus  the curve  $N_c^{\text{FD}}$, in
 its whole, should  not be taken too much   seriously.  However if  we
 take, as estimates of the error-bars  associated to this curve, those
 provided by Calabrese
\etal in their computation \cite{calabrese04} one can safely trust in
a finite  portion of the part {\sl  below} S.} that gathers the curves
$N_c(d)$ ---  the critical value  of the  number  of components  above
which  the transition is predicted to  be of second order --- obtained
from the different RG approaches  (the perturbative ones and also that
obtained  with  the NPRG).   Fig.\ref{courbes_ncd} is,  in particular,
very   symptomatic of the  existence   of  a  discrepancy between  the
different {\sl perturbative}   approaches since one clearly  sees that
the curves  $N_c^\epsilon$ (obtained within the minimal substraction ($\overline{\hbox{MS}}$)  scheme
{\sl with an $\epsilon$-expansion} \cite{calabrese03c}) and $N_c^{\text{FD}}$
(obtained  within   the   $\overline{\hbox{MS}}$  scheme  {\sl without
$\epsilon$-expansion} \cite{calabrese04}) are  {\sl incompatible}  for $N\lesssim 6$
and this, {\sl independently} of  the results obtained within the NPRG
approach.

In this respect, we are a little bit surprised that the authors of
\cite{pelissetto06}, in  their  Comment, do  {\sl never}  mention this
discrepancy and not  even the curve  $N_c^\epsilon$ (that  have been obtained
from a {\sl five-loop} computation  \cite{calabrese03c}) displayed in Fig.\ref{courbes_ncd}
from which originates an important part  of the controversy.  Instead,
they focus  on the ``{\sl  difference between the perturbative results
and those obtained  by  using the [\dots]  FRG}''  \cite{pelissetto06}
that is not our purpose.

Implicitly in their Comment, and more explicitly in their article
\cite{calabrese04}   the origin
of  the problem that  we raise  would come from   the inability of the
$\epsilon$-expansion to provide   a   correct   description of  the     three
dimensional physics. In contrast to the authors of \cite{pelissetto06}
we  think  that  the agreement between   the   curves $N_c^{\epsilon}(d)$ and
$N_c^{\text{NPRG}}(d)$  ---    that  are obtained    from  drastically
different computations --- is  rather  remarkable and leads to   trust
both the $\epsilon$-expansion and the NPRG approaches.  Also, the FD approach
(including calculations  in  the massive zero-momentum (MZM) scheme  or  $\overline{\hbox{MS}}$
scheme without $\epsilon$-expansion) appears to be very isolated as it is the
{\sl only one}  leading to the prediction  of criticality in $d=3$  in
the Heisenberg and XY cases. This is precisely this fact that leads us
to  search  for ---  and actually  to find  --- a  flaw   in the FD
approach.

It  is    now time  to  answer in   detail to  the   points  raised in
\cite{pelissetto06}.

\vspace{0.3cm} 

\section{\large II- Detailed answers to the Comment}

{\bf 1)}  According to the   authors of \cite{pelissetto06}  there  is
``{\sl no theoretical justification}'' for the requirement that ``{\sl
a physical FP of a given Hamiltonian  must survive up to $d=4$ [\dots]
and become the Gaussian FP in this limit}''.   We shall comment on the
word  ``physical''   in point   {\bf  2)}  below.  We start    here by
discussing the question of the ``survival'' of a  FP as a Gaussian one
in $d=4$.   We  strongly disagree with  the  statement that  there  is
``{\sl   no theoretical justification}''  for  this.  Indeed, in Field
Theory    as  well as in  Statistical   Mechanics,  the  choice  of a
Hamiltonian implies a  choice  of the  most relevant operators  of the
theory. The common belief  which, as far as  we know, is confirmed  by
all  well-controlled situations, is that   this choice relies on {(i)}
the existence of an upper  critical dimension, { (ii)} the possibility
of a ``naive'' power counting  performed around a Gaussian fixed point
(FP) and,  therefore, { (iii)}  the fact that   the theory is infrared
free  in this  dimension.   In  the  case where   there exists,  in  a
dimension $d=4-\epsilon$,  a non trivial  FP, whose coordinates  are of order
$\epsilon$, the flow is well controlled everywhere  between the UV and the IR
scales.   When this    FP can  be extrapolated  down    to $d=3$,  the
perturbative FD approaches performed directly in $d=3$ are expected to
safely describe the   physics in this dimension  although  they do not
refer explicitly to the upper  critical dimension.  Moreover, still in
this  case, there are strong indications  that the field theory can be
``constructed''  directly in $d=3$ \cite{parisi80}.  This is,  in particular, the case
of  the $(\phi^2)^2$   theory where the  $\epsilon$-expansion  predicts  a FP in
$d=3$, which validates the MZM perturbative scheme.  Thus, at least in
all well-controlled situations, and contrary to what is asserted in
\cite{pelissetto06}, there   is a  link  between the  FD approaches in
$d=3$  and the $\epsilon$-expansion and,  thus, an implicit link between this
FD approach  and the Gaussian behaviour of  the critical theory in the
upper critical dimension.

Now consider a  theory  where  a FP   found in  $d=3$  is  related  by
continuity to a non-Gaussian FP in $d=4$.  (This is what we have found
in the  FD perturbative  approach for  frustrated magnets for   the FP
$C^+$ when --- and {\sl only when} --- $N\lesssim 6$ and for the FP $P$ found
in the cubic model for $v<0$.)  If this turned out to be true, and not
an artefact of  the FD approach (whereas  we think it is),  this would
mean that  the link between the  $\epsilon$-expansion and FD approaches would
be broken. Thus the theory would have no  more a well identified upper
critical  dimension and it would  be {\sl non  trivial} in $d=4$.  For
$(\phi^2)^2$-like theories describing the frustrated and cubic cases this
would be completely  new, unexpected  and, if true,  would be  of {\sl
utmost} interest !

As   a conclusion, we {\sl  do}  think that  there  is an implicit but
strong  link  between  the   trivial  IR  behaviour of  $(\phi^2)^2$-like
theories in $d=4$ and the validity  of the description of the critical
physics in $d=3$ by means of FD approaches.
 
\vspace{0.3cm}


{\bf 2)} In  Comment  \cite{pelissetto06}, the words  ``physical'' and
``survive''  in  the   sentence:  ``{\sl a  physical    FP of a  given
Hamiltonian must survive up to  $d=4$ [\dots] and become the  Gaussian
FP in this limit}''  are clearly, from what follows, understood by
the  authors  as ``having  real   coordinates''.  This leads  them  to
conclude that    ``{\sl according  to  this criterion,   all FPs  with
$N\lesssim21.8$ should  be  considered as  spurious }'' and  that ``{\sl this
condition  is  very restrictive  and contradicts several well-accepted
theoretical  results}''.  This, by  no  ways,  corresponds to  what is
written in our article. Indeed our criterion   to consider   a FP
either as  physically   relevant or as  spurious,  is based   on an
analysis of the roots of the $\beta$ functions that  can be either real or
complex. We  have explicitly written   in foonote [27]  of our article
that ``{\sl If one follows $C_+^{\text{FD}}$  along a path starting in
$d=3$, going to $d=4$ and crossing $N_c^{\text{FD}}(d)$ above $S$, its
coordinates  become complex in  $d=d_c(N)$ and  go  to zero for  $d=4$
where it is thus  the Gaussian FP}''.   Our criterion does not consist
in rejecting  FPs whose coordinates become  complex when the dimension
$d$ is increased but in rejecting those that are {\sl non-Gaussian} in
$d=4$. As a consequence all FPs in $d=3$ having $N\gtrsim  6$ (that is those
``above  the  singularity  $S$'') are,   according to  us,  physically
relevant; and vice versa.  This is,  again, explicitly written in page
3, column 2, where we say that, in  the frustrated case, the fact that
the FP are  not Gaussian ``{\sl happens for  all  values of $N\lesssim  6$}''
and, by no means, for all values of $N\lesssim21.8$ !

As a conclusion this part of Comment \cite{pelissetto06} is based on a
trivial misunderstanding of our article.  Our criterion is by no means
``{\sl very restrictive}'' and, contrary to what is asserted in
\cite{pelissetto06}, agrees with all ``{\sl well-accepted
theoretical results}''.


\vspace{0.3cm} 

{\bf 3)} According to   the authors of \cite{pelissetto06}  ``{\sl  We
must observe   that   the perturbative  results  of   Ref.[3]  find no
difference between  the FPs with  $N\gtrsim  6$ [\dots]  and  those with $N\lesssim
6$}''.   This statement    is  {\sl incorrect}  since   it  ignores an
important   fact   that   has   been   missed   by    Calabrese  \etal
\cite{calabrese04}, {\it i.e.}  the  existence of a singularity $S$ in
the $(N,d)$ plane which makes the coordinates of the FP $C_+$, $u_1^\star$
and $u_2^\star$, multivalued functions of $N$ and $d$, as explained in our
footnote [27] of
\cite{delamotte06}.  This fact is manifest when one follows the FP $C_+$
 by continuity along  a path encircling the  singularity $S$: after  a
 round trip, the coordinates of the FP thus obtained have changed. The
 existence of such  a singularity is also  at  the origin of the  fact
 that  the FPs  obtained  in $d=3$ are or   are not  Gaussian in $d=4$
 depending on whether the path  followed to  reach the upper  critical
 dimension   passes above or   below  the singularity $S$.  Therefore,
 contrary to the statement of the authors of \cite{pelissetto06} there
 {\sl is} a {\sl  fundamental} difference between  the cases $N\lesssim6$ and
 $N\gtrsim 6$.


\vspace{0.3cm}

{\bf 4)} According  to the authors  of \cite{pelissetto06} ``{\sl  the
difference  between the perturbative    results and those  obtained by
using  the functional  renormalization  group   (FRG) [5-7]  is   only
quantitative [\dots].  But there  are no conceptual differences as the
authors [of this reply] apparently imply.}''

First we, again, emphasize as in the  Introduction to this Reply, that
the   purpose of our  article  is  not  to  oppose nonperturbative and
perturbative  approaches.  It is  to  try to understand  why two  {\sl
perturbative} approaches  relying on  the  {\sl  same} renormalization
scheme ($\overline{\hbox{MS}}$ scheme) and differing {\sl only} by the
way of solving  the FP equations lead to  {\sl qualitatively}, and not
only quantitatively, different results. 

Second, concerning the  nature 
  of  the  difference  between  the  FD
approach and the $\epsilon$-expansion, we disagree with the authors of
\cite{pelissetto06}. The very questions we have addressed 
 are { (i)} to know whether there exists a FP in $d=3$ for $N=2$ and 3
 { (ii)} whether this  FP is non-Gaussian  when followed in $d=4$ and,
 finally, at   a more technical  level,   { (iii)}  does  there  exists a
 singularity $S$ in the coordinates $u_1^\star$ and $u_2^\star$ of $C_+$ taken
 as functions of  $N$ and $d$.  To  all these questions the  answer is
 {\sl positive}  in the $\overline{\hbox{MS}}$ scheme approach without
 $\epsilon$-expansion whereas it is  {\sl negative} in the $\epsilon$-expansion (and
 in the NPRG) approaches.

We think that  these discrepancies are, indeed,  ``{\sl conceptual}''.
Moreover, we believe that  we have identified  the very origin of this
``{\sl     conceptual}'' difference:  the    FP  identified within the
$\overline{\hbox{MS}}$ scheme without  $\epsilon$-expansion corresponds  to a
{\sl spurious} solution  of the FP  equations that are solved at fixed
$d$.


\vspace{0.3cm}

{\bf  5)} According   to   \cite{pelissetto06}  ``{\sl  The  behaviour
observed here is analogous to that  found in the Ginzburg-Landau model
of superconductors [\dots]   The criterion proposed  in  Ref.[1] would
thus predict a  first-order  transition for the physical  case  $N=1$,
contradicting  experiments   [9], and  also  general duality arguments
[10], FRG calculations [11], and Monte Carlo simulations [12].}''  The
answer given in point  {\bf 2)} above  also applies to this  case: our
criterion would  {\sl  obviously not} lead  to  predict  a first-order
transition in superconductors since   our criterion has nothing  to do
with  the critical   value of  $N$  in  the upper critical   dimension
$d=4$. In particular, we  emphasize that our considerations absolutely
do not exclude the existence of a FP for $N$ smaller than $N_c$.

\vspace{0.3cm}


{\bf 6)} The authors of
\cite{pelissetto06} state that  ``{\sl There is another condition that is crucial: 
the  three-dimensional FP must be  connected  by the three-dimensional
renormalization-group  flow to the Gaussian  FP [4,15]. If this is the
case, at  least for the  massive zero-momentum  (MZM)  scheme, one can
give   a  rigorous nonperturbative  definition  of the renormalization
group flow and  of all quantities   that are computed  in perturbation
theory}''. They also state that ``{\sl In a well-defined limit [\dots]
long-range quantities [\dots] have the  same perturbative expansion as
the corresponding  quantities in  the  continuum  theory  for the  MZM
scheme}''  and   that   therefore    ``{\sl everything    is   defined
nonperturbatively and rigorously in three  dimensions and there is  no
need of invoking the existence of a four-dimensional FP }''.  We agree
with all these statements (that we have already partially discussed in
point {\bf    1)}).   However,  there is  a    fundamental  assumption
underlying all these statements:   they are valid {\it provided  there
exists an  IR stable FP in $d=3$  \dots which is precisely the dubious
point !}

As a   conclusion,  all the arguments   invoked in \cite{pelissetto06}
about  the nonperturbative and rigourous  definition of the MZM scheme
approach, as well as  the claim that   the behaviour of the  theory in
$d=4$ would not be  relevant for the  three dimensional physics, would
be of interest  \dots if there were  no  controversy  about the critical
behaviour of the frustrated systems in $d=3$.

\vspace{0.3cm}

{\bf 7)} According   to \cite{pelissetto06}  our statement that    for
$N=2,3$ one  can follow the  FP $C^+$ up to  $d=4$ is ``{\sl incorrect
and  is based     on  an  incorrect  use  of   the   conformal-mapping
method}''. We completely agree with, and are aware  of, what is stated
in \cite{pelissetto06}  about limitations of the resummation procedure
used in \cite{delamotte06}.  Concerning this  point we have written in
our article  that when one follows the  FPs from $d=3$ to $d=4$ ``{\sl
the  FPs  $P$   and   $C_+^{\text{FD}}$  lie out  of   the  region  of
Borel-summability  in  $d=4$.    Thus their   coordinates   cannot  be
determined   accurately}''.   However this   point  is {\it completely
irrelevant} to the question raised in  our article: the identification
of   the (non-)Gaussian  character  of  the  FP    in $d=4$. And   the
non-Gaussian  character  of  the FPs with   $N\lesssim6$ at $d=4$ is  {\sl doubtless}.
Indeed, let us suppose on the contrary that a FP, with $N\lesssim6$, followed
from $d=3$ to  $d=4$ is, actually,  a Gaussian one  in $d=4$. In  this
case,   its coordinates just  below  this dimension would be extremely
small  and  thus it  could obviously  be  obtained within perturbation
theory without any resummation procedure. Thus, as we clearly state in
our article,  our procedure is completely valid  to decide whether the
FP  is  or  is  not  Gaussian in  $d=4$   (although not  sufficient to
determine its coordinates if it is not Gaussian, what we do not mind anyway).

We add, as we  have already emphasized above, that  the occurence of a
non-Gaussian FP in   $d=4$ is deeply related  to  the existence  of  a
singularity $S$ in the $(N,d)$ plane.  An  important fact is that this
singularity $S$ lies either {\sl inside or  just on the border of} the
region of   Borel-summability.   Its  existence is    thus  doubtless,
according to the standards of \cite{pelissetto06}.

As   a   conclusion  the  arguments    raised   by   the  authors   of
\cite{pelissetto06} concerning our use of the conformal-mapping method
are completely irrelevant for our purpose.

\vspace{0.3cm} 

{\bf 8)} Concerning numerical results, according to the authors of
\cite{pelissetto06} ``{\sl All numerical  and experimental results are
consistent with the  predictions of perturbative field theory}'' since
``{\sl the existence  of a stable FP  does not imply that all  systems
with   the    given     symmetry   undergo   a    second-order   phase
transition.}''. We --- obviously --- agree with the last argument that
the existence of   a FP in  a field  theory  does not  imply that  all
systems undergo a second order  phase transition.  However we disagree
with the  conclusion (that the numerical  and experimental results are
consistent with perturbative theory) which is drawn from it. Indeed:

1) While the first  order behaviour does  not  contradict the existence of a
FP one could expect,  from  the existence of such  a  FP, a basin   of
attraction  of   finite extension and, thus,   that  some materials or
numerically simulated systems   exhibit a second order behaviour  with
the predicted critical exponents.  This is  not the case apart from an
isolated simulation performed   on a  lattice  discretization of   the
Ginzburg-Landau  Hamiltonian that apparently  leads to  a second order
behaviour
\cite{calabrese04}.  We say ``apparently'' since we have been used to
claims of the  existence of second order  behaviours for systems  that
have   been  subsequently discovered  to   undergo a  weak first order
transition.  (As an example, stacked triangular antiferromagnets (STA)
have long been thought  to undergo continuous transitions until larger
system sizes  have been considered.)  This  could also be the case for
the simulation performed in
\cite{calabrese04}.

2)  The  numerical simulations leading  to a   (apparent) second order
behaviour  (it is now  recognized that most  of them are of weak first
order) were considered in ``{\sl substancial agreement}''
\cite{pelissetto01a} with the six-loop computation performed in the
MZM scheme \cite{pelissetto01a}, thus giving a credit to the existence
of the FP identified  in this way.  But, strangely, the fact that more
accurate simulations eventually found  first  order instead of  second
order transitions has never been considered by the authors of
\cite{pelissetto06} and   \cite{pelissetto01a}   as  contradicting 
  the results of  the FD approach.  This means  that  the existence of
  second  or first order  phase transitions  {\sl equally} confirm the
  predictions of this approach !

3) Rather than  focusing on  the first order behaviour   in  general it is
more instructive  to address the   question of the  occurence of  {\it
weak} first order behaviour.   Indeed, in presence  of a standard  FP
characterizing a second order  phase  transition, one can expect  weak
first order behaviour.  However this can happen only for very special
initial conditions of the RG  flow such that the point  representative
of the system  in the space of couplings  is in the  runaway
region --- leading to first order behaviour  --- but very close to the
boundary between the first  and second order regions  so that the flow
is slow  and produces a very  large correlation length. In a numerical
investigation of  a whole  family of  frustrated magnets  it has  been
shown by A.  Peles, B.W. Southern and some of the present authors
\cite{peles04,bekhechi06} that, actually, weak first  order transitions  occur {\sl
generically}  in  these systems.   As these  authors have argued, this
contradicts the usual interpretation given for the ``occasional'' weak
first order behaviour described above.  Thus there must exist another
explanation to these generic   weak first order behaviour.  The  NPRG
approach  \cite{delamotte03} provides such   an explanation:  it shows
that the weak first order  behaviour  that occurs  in frustrated magnets
does  not rely on the usual explanation above but rather on the existence
of a  generic slow  flow  (than occurs  even in  absence  of a FP).
Thus, contrary to what the authors of
\cite{pelissetto06} say the ``{\sl quoted results [17-19]}'' are not
irrelevant for the discussion. On the contrary: {  (i)} they show that
frustrated systems  initially thought to undergo  a second order phase
transition actually have first order behaviour  { (ii)} they show that
this first order behaviour  is {\sl generic} {(iii)}  they point out a
weakness of  the perturbative FD  approaches that are {\sl  unable} to
explain the existence of {\sl generic} weak first order behaviour.

Finally the  authors of  \cite{pelissetto06}   argue that ``{\sl   the
results of Ref.[3] and of Ref.[20]  - they find continuous transitions
for $N=2$ and  $N=3$,  respectively  - are  only  consistent with  the
presence of a  stable FP  and thus  do not   support the scenario   of
Ref. [1]}''.  We recall   that the past experience   in the domain  of
numerical  simulations   of   frustrated magnets  has    been  largely
controversial (see \cite{delamotte03}   for instance).   Also in  most
cases improving the  method of analysis  (for instance  by using Monte
Carlo Renormalization Group methods
\cite{Itakura01} or dynamical methods  \cite{bekhechi06}) has  lead to
the conclusion of a {\sl weak first order} transitions.

\vspace{0.3cm} 

{\bf   9)} Concerning the experimental   situation,  according to  the
authors of \cite{pelissetto06}, the  ``{\sl results of  Ref.[21] cited
in   Ref.[1] are perfectly consistent   with  perturbation theory.  We
discussed  in  detail  easy-axis systems   in Ref.[22] and  showed two
possible phase diagrams  compatible with perturbation theory}''.  Here
the authors refer to their own work with P. Calabrese on multicritical
behaviour in frustrated  systems \cite{calabrese05} that would explain
the first order behaviour found in CsNiCl$_3$
\cite{quirion06}. However, according to the 
authors of  \cite{calabrese05} (abstract of  this article): ``{\sl the
transition  at the  multicritical  point  is   expected to be   either
continuous and controlled by the $O(2)\otimes O(3)$  fixed point or to be of
first  order}''.  They also add in  their Comment that:  ``{\sl Due to
the `focus'-like nature of the FP [23] the approach to criticality may
be quite complex.  Effective exponents may even change nonmonotically
\dots}''.  As in point  {\bf 8)} above,  such statements make any experimental 
(and numerical) behaviour,  of first or second  order with any set of
critical exponents, to  be  compatible with perturbation theory !

Concerning easy-plane systems, according to the authors of
\cite{pelissetto06}, ``{\sl all experiments observe continuous
transitions,  and  thus  they  are  compatible with  the  perturbative
results}''. We  emphasize here that there  is  no definitive statement
about the  transition in these systems.   On the contrary, some of the
present  authors have shown \cite{delamotte03}  that  several facts go
against the belief  that the  transitions  are  of second  order:  the
critical  exponents found are   {\sl non universal},  scaling laws are
{\sl violated}, the  anomalous dimension --- or   exponent  $\eta$  ---  is
{\sl negative}, etc.  It is absolutely  not excluded  that,  as in the
case of CsNiCl$_3$ and in almost all numerically simulated models that
were   believed   to undergo  a second   order  phase  transition, all
easy-plane systems will   be finally claimed  to   undergo first order
transitions.

\vspace{0.3cm} 

{\bf 10)} Concerning the cubic model, the authors of
\cite{pelissetto06} contest its use since the situation would not be, 
in   this case, ``{\sl   well   established}'',  contrary to   what we
claim. To support this  statement they invoke --- in  this case too  !
--- the possible failure of the $\epsilon$-expansion in the region $v<0$ that
we precisely investigate in our article.

1) To our knowledge this is the {\sl  first time} that  the use of the
$\epsilon$-expansion is questioned in  the context of  the study of the cubic
model.  On the   contrary,  all approches  used to   investigate,  for
instance, the critical value $N_c'$ (not to be confused with the $N_c$
of  frustrated  magnets)  above  which  the  cubic  FP  is  stable, do
coincide.   For  instance,  according  to  an article \cite{carmona00}
gathering J.M. Carmona  and the authors  of \cite{pelissetto06}, it is
found  that    $N_c'=2.89(4)$ from a  six-loop   perturbative approach
performed  in $d=3$ and  $N_c=2.87(5)$  from a five-loop $\epsilon$-expansion
approach.

2) According to the authors of
\cite{pelissetto06}  something special should happen  in the case $v<0$. 
Indeed  according to   them the critical   behaviour of  :  ``{\sl the
antiferromagnet  four-state Potts model   on a  cubic lattice  [26-28]
[\dots] should be described by the  $N=3$ cubic model with $v<0$ [29]}
''. They   add that  ``{\sl Contrary  to  the  claim  of Ref.[1],  all
numerical results are    consistent with a continuous  transition:  at
present there is no evidence of first order transitions}''.

While we did not write anything in our  article about the behaviour of
the antiferromagnetic four-state  Potts model we, however,  completely
disagree with  the statement  that  ``{\sl all numerical  results  are
consistent  with a continuous  transition}''  in these systems.  It is
true that some simulations
\cite{ueno89} have lead to the claim of a second order
behaviour for  these systems.   However it  has been  recognized  with
further investigations that this conclusion was hasty since
\cite{itakura99}:    ``{\sl   the     Hamiltonian  for    the    $q=4$
antiferromagnetic Potts model on both simple cubic and body-centered-cubic lattice
 is far apart any
fixed  point  and  the  largest    simulated  size  $L=96$  is   still
insufficient to extract asymptotic critical behavior. However, we have
found    that the Hamiltonian   moves towards  the strong $\langle100\rangle$-type
anisotropy (large negative $v$) direction as it is renormalized. Since
the recent  field-theoretical$^{12}$  and  Monte Carlo$^{14}$  studies
indicate the absence of RG fixed point in  the $v<0$ region, we expect
that the transtion is  a first-order one.}''  We  do not see here what
is consistent with a continuous transition.

Also we note  that the authors of the  Comment  themselves, with their
collaborators, claim in Ref. \cite{calabrese02c}  (P2) that ``{\sl the
four-state [antiferromagnetic  Potts] model  is   expected to show   a
first-order transition}'' and (P5) that ``{\sl In the four-state case,
the weak first  order transition expected in  the pure case should not
be softened by random dilution}''.

 As for the cubic model itself in
\cite{carmona00} (footnote [10]) they claim that ``{\sl A high-temperature
 analysis on  the   fcc lattice indicates  that  these  models have  a
 first-order transition for $N>2.35 \mp 0.20$.   This is consistent with
 our argument  that predicts the transition  to be of  first order for
 any $v_0<0$ and $N>N_c$. More  general models that have Eq.(1.1) [the
 cubic  model] as their continuous  spin limit  for $v_0\to-\infty$ have also
 been considered in Ref.12.  The  first-order nature of the transition
 for negative (small) $v_0$ and  large $N$ has  also been confirmed in
 Ref.11.}''  Also in \cite{calabrese03d}, where a six-loop computation
 has been performed, they claim (P1) that ``{\sl  for $w<0$ [$w$ being
 the coupling of the cubic term],  the RG flow  runs away to infinity,
 and the   corresponding   system  is  expected   to  undergo  a  weak
 first-order transition}''.  Again  we do  not see any   controversial
 situation  here: {everything  seems  to  favor a  first  order  phase
 transition    contrary      to  what    is     claimed   in   Comment
 \cite{pelissetto06}.

While we acknowledge   that  the description  of  the  antiferromagnetic 
four-state   Potts model could  be  problematic, it certainly does not
question the structure of  the flow diagram of the  cubic model in the
region $v<0$ as the  authors of \cite{pelissetto06} suggest,  at least
in their Comment.

\vspace{0.3cm}

{\bf 11)} Still on the cubic model, according to the authors of
\cite{pelissetto06} ``{\sl the analysis of the perturbative series
 does not provide compelling evidence for the existence of a new FP in
 cubic models with  $v<0$  and $N=3$}''.  To   support this claim  the
 authors of \cite{pelissetto06} have  repeated   our analysis on   the
 cubic model with the help of a  FD analysis and have  shown that only one
 half   of the different   resummations   leads  to   a FP. Thus   our
 ``demonstration'' would  not  be  convincing.   Also  comparing their
 percentage of  FPs  obtained  in the cubic case to that
 obtained in the  frustrated  case, the authors of  \cite{pelissetto06}
 contest our  statement  that  the   cubic model   show  ``{\sl
 similar convergence properties}''  to    the frustrated one.

First, we note that  the authors of \cite{pelissetto06} have confirmed
an  important  result of   our article which    is that FD  approaches
generate dubious FPs that are not obtained within the $\epsilon$-expansion.

Second,   the criterion of     the authors of  \cite{pelissetto06} for
accepting  or rejecting  a  FP is  very vague.  They
consider  ranges of values of the   resummation parameters ($-1\leq \alpha\leq 5$
and $2\leq b\leq    20$)  that are  completely  arbitrary.    For  instance,
by considering larger ranges  of values of $\alpha$ and/or $b$,  they could obtain  a 
percentage of rejections as small as wanted.

Third, contrary to what is claimed by the authors of
\cite{pelissetto06}, our statement that the cubic-model results show
``{\sl similar convergence properties}'' as in  the frustrated case is
largely justified. Indeed, we have  extensively studied the properties
of convergence of the exponents $\omega$ and $\nu$ at the FP using criterions
of best apparent convergence or  principle of minimal sensitivity.  We
have been lead  to the conclusion that  there exists an  error between
the four and five loop  results that is of  order 40 $\%$ in {\sl both}
frustrated and  cubic cases. Our study,  based on these criterions, is
far  more instructive that   statistics performed varying the
resummation parameters on arbitrary  domains that does not provide any
reliable information on the nature (spurious or  physical) of the FP.

Finally concerning the sentence ``{\sl The  difference between the two
cases [cubic   and frustrated  ]  [\dots] are    so evident,  that  no
additional comment is  needed !}''.  We have no  problem to admit that
the cubic FP could display weaker stability properties with respect to
variations of the resummation parameters than the frustrated one, what
remains  to be proven  using other  criterions than those of \cite{pelissetto06}.
 However, this
is not a relevant point for our  purpose.  Indeed, there is absolutely
no reason that the  spurious FPs  obtained  in  two  different models
display the same  (quantitative)  behaviour. The relevant point  is  a
comparison to the Ising, or more generally $O(N)$, model for which, at
the same order, one  has an error  one hundred  times smaller  than in
both the frustrated and cubic cases!  As the authors of
\cite{pelissetto06}  say ``{\sl  Differences are  so  evident, that no
additional comment is needed !}''.

To conclude,  our study of the  cubic model is  appropriate to
demonstrate the  spurious character of  the FP found in the frustrated
case.   Indeed, {(i)}  it is largely  less controversial  than  the
frustrated case { (ii) } it admits  a FP that  very probably is 
 a spurious one { (iii)} the properties of the convergence of the
physical  quantities at this  FP are very close to  those found in the
frustrated case if one adopts criterions { based on optimization of
the results.

\vspace{0.3cm} 

{\bf 12)} The authors of  \cite{pelissetto06} propose a way of  ``{\sl
Reconciling the   different  approaches}''.   According to   them  the
differences observed between  the  perturbative FD and  FRG approaches
would be related to the ``{\sl crudeness of the approximations used in
Refs.[5,7] [the FRG approach]}''.

First   we note,  again, that  these   authors focus on the opposition
between perturbative and nonperturbative   methods, which is  not  our
true motivation.  However let us {\it  imagine} that these authors are
right.    Then, how to explain    the  agreement between the five-loop
$\epsilon$-expansion  (that is certainly not  a crude approximation according
to the standard) and FRG approach ?

Second,  we definitively reject  the argument of  crudeness of the FRG
computations. Indeed,   as explained at length  in \cite{delamotte03},
the FRG analysis  of  the frustrated  magnet  succeeds several  tests:
agreement with the results obtained within a low-temperature expansion
of the nonlinear  sigma  model around two  dimensions \cite{azaria90},
agreement  with  the  weak-coupling  expansion of  the Ginzburg-Landau
model around four dimensions
\cite{jones76,bailin77},
agreement  (better than 1 $\%$)  with the large-N results (performed up
to   order ${1/N^2}$) \cite{bailin77,kawamura88,pelissetto01b} in  any
dimensions between 2 and 4 dimensions, agreement  with the $N=6$ Monte
Carlo   results   \cite{loison00},  agreement   with the $\epsilon$-expansion
performed at five-loops everywhere between 2 and 4 dimensions
\cite{calabrese03c}.  Finally, the  stability   of the   results  with
respect to changes of the field content has been also checked by using
more refined truncations.

One cannot say  that   the same  checks  have  been performed  for  FD
computations.   In  fact, there is  even   a  discrepancy between  the
different FD results if one considers the MZM scheme in which no FP is
found  between  $N=5$  and  $N=7$ (while  we recall   that  it is well
established from  Monte Carlo simulation  that there is a second order
phase  transition  in the  $N=6$ case).   Also, as said  above, if one
considers the numerical and experimental  situations there is only one
case, a Monte Carlo simulation
\cite{calabrese04},  where a   second  order transition with    critical
exponents  close  to    the perturbative   FD  predictions  have  been
found. However, looking at  the most recent experiments and  numerical
simulations one finds  more   and more systems where  the  transition,
considered in the past as a second  order transition, is discovered to
be,  in fact,  of  weak first  order, in   contradiction  with the  FD
perturbative approaches.

\vspace{0.3cm} 

\section{\large III- Conclusion}

The points raised  by  the authors of \cite{pelissetto06}  come  from
either misunderstanding of our article or from  unfounded assumptions, as
we have shown here. Also,   contrary to what is   claimed in the  Comment
\cite{pelissetto06},  our  results   agree  with all
existing and   well established theoretical  or  experimental results.
Moreover, we maintain to have provided a solution {\sl explaining} the
manifest contradiction between the $\epsilon$-expansion (that agrees with the
NPRG approach) and  the   FD perturbative approaches that   predict  an
unobserved second order  critical behaviour both in frustrated magnets
and in the cubic model.

\end{document}